\documentstyle[11pt,fleqn,epsf]{article}
\def\ref{par\noindent\hangindent=6mm\hangafter=1}
\begin{document}
\baselineskip 8mm

\begin{center}
{\bf Note on the Demkov-Ostrovsky nodeless sector}

\bigskip

 H. Rosu\footnote{E-mail: rosu@ifug3.ugto.mx}    
 and J. Socorro\footnote{E-mail: socorro@ifug4.ugto.mx}

{\it Instituto de F\'{\i}sica de la Universidad de Guanajuato,
Apdo Postal E-143, Le\'on, Guanajuato, M\'exico}

\end{center}

\bigskip

{\bf Summary.} - We briefly tackle the following concepts in the
Demkov-Ostrovsky (DO) nodeless sector:
(i) orbital impedance, (ii) orbital capacity,
(iii) closeness to reflectionlessness.
Moreover, using previous supersymmetric results for the DO problem,
a strictly isospectral effect in the DO orbital impedances is discussed
and explicit plots are displayed for the Maxwell fisheye lens.
This effect, though rather small,
is general, that is, it may apply to any focusing structure.





Because of one of the optical-mechanical analogies, the DO
problem can be stated as a Schr\"odinger radial (half line) equation at zero
energy with the focusing potential
$
V_{\kappa}(\rho)=-w/\rho ^2[\rho ^{-\kappa}+\rho ^{\kappa}]^2
$,
where $w>0$ is a (Sturmian type) coupling constant, $\rho$ is a scaled radial
variable ($r/R$),
and $\kappa$ ($> 0$) is the Lenz-Demkov-Ostrovsky parameter, which is
unity for
the Maxwell fish eye (MFE) lens, and one half for an atomic aufbau
model \cite{do}.
In previous works \cite{dos}, a supersymmetric, Witten approach of
the DO problem in the nodeless radial sector $n=l+1$ (1s, 2p, 3d ...)
has been put forward.
Moreover, the strictly isospectral double Darboux method has been worked out
with the interesting result that a one-parameter family of Maxwell lenses
having the same optical scattering properties in the nodeless sector
might exist \cite{dosr}.

In this note, still within the same nodeless sector,
we will refer to the concepts of DO orbital impedance
and orbital capacity. Moreover, we mention a problem
that we call ``closeness to reflectionlessness". The strictly isospectral
effect \cite{dosr} is graphically displayed using the language of impedances
that may have some advantages not only in technological applications.

\bigskip

{\em (i) Orbital impedance:}
Martin and Sabatier
showed in a general context that a localized, strictly positive zero energy
solution (zero mode) of a Schr\"odinger equation can be interpreted as an
inverse impedance \cite{ms}.
If we claim that the inverses of the DO radial factors in
the nodeless sector, i.e.,
$
f_{\kappa ,l}^{-1}\equiv Z_{\kappa ,l}
=\rho ^{-(l+1)}(1+\rho ^{2\kappa})^{(2l+1)/2\kappa}
$
are  orbital ``impedances", then these DO
impedances are singular ($\propto 1/\rho^{l+1}$) at the center of
the DO spheres, and also increase sharply beginning at an outside distance
of about one diameter, with no relevant structure.
On the other hand, we have found that the strictly isospectral
radial factors obtained by the double Darboux method \cite{dosr} are
more interesting, precisely because they have a more intricate
spatial behavior with the isospectral Darboux parameter playing an important
role in the problem. Thus, it may be indeed helpful
to consider the inverses of the DO isospectral radial factors as isospectral
orbital impedances, having the form $Z_{iso}\equiv
f_{\kappa ,l}^{-1}(I_{\kappa ;l}+\lambda)$, where
$I_{\kappa ;l}=\int_{0}^{\rho}f^{2}_{\kappa,l}(\rho ^{'})d\rho ^{'}$ and
$\lambda \in (0,\infty)$ is the isospectral family parameter, i.e.,
mathematically speaking, the Riccati integration constant. $Z_{iso}$ is also
singular at the DO center. Therefore, in order to show in a clear-cut way the
strictly isospectral effect for the MFE lens, we did plots of the ratio
${\cal R} _{l}(\rho)=Z_{iso}/Z_{1,l}=
(I_{1,l}(\rho)+\lambda)$
for the partial waves from $l=1$ to $l=3$ and for $\lambda =0.01,\;
0.1,\;1,\;10$
in each $l$ case (see the plots). As a matter of fact,
the integrals $I_{1,l}$ can be calculated analytically using the recurrence
formula $\int\frac{x^{m}}{(1+x^2)^{n}}=\int \frac{x^{m-2}}{(1+x^2)^{n-1}}-
\int \frac{x^{m-2}}{(1+x^2)^{n}}$, $m=2(l+1)$ and $n=2l+1$.
The effect is stable and diminishes substantialy when
one goes from one $l$ to the next $l+1$ one. It starts to reveal itself at 
about half of the radius $R$ within the focusing structure, reaches a
maximum  at $\leq \sqrt{2}R$ outside the lens which is moving toward the
surface at higher $l$, and is insignificant beyond about four $R$s.
These strictly isospectral effects may show up in any focusing structure,
for example in radar applications of the MFE lens.
The isospectral parameter $\lambda$ is controling the scale between the
isospectral impedance
and the normal one. This is why, we have chosen four scales corresponding to
$Z_{iso}$ one hundred times smaller than $Z_{1,l}$,
ten times smaller, equal, and ten times bigger
than
$Z_{1,l}$ for $\lambda=0.01$, $0.1$, $1$, and $10$, respectively.
The relative importance
of the strictly isospectral effects depends on the value of $\lambda$.
The case $l=0$ is 
special in the sense that ${\cal R} _{0}(\rho)=
\rho -{\rm arctg}\rho +\lambda$ is a steadily increasing function and
therefore does not show a peak structure as the other partial waves do.
One can calculate easily the position of isospectral impedance peak to be
$\rho _{M}=\sqrt{(1+\frac{1}{l})}$, which in the limit of high $l$ almost
reaches
the surface of the MFE lens, and also the velocity of this peak structure
in the $l$- space $v_{M}=-\frac{1}{2l\sqrt{l(l+1)}}$, showing that in the
limit of high $l$ the peak is almost static.

\bigskip

{\em (ii) Orbital capacity:}
Suppose we adopt a thermodynamic view consisting in
considering the ``fermionic" (scattering)
effective potential $U^{+}(l)$ as a kind of metastable free
energy and thus we introduce the DO ``orbital capacity"
$C^{+}_{l}=\frac{dU^{+}}{dl}$, where
$U^{+}(l)=l(l-1)/\rho ^{2}-(2l+1)(2l-2\kappa-1)/
[\rho ^{2(1-\kappa)}(1+\rho ^{2\kappa})^{2}]+
2(2l+1)/[\rho ^2(1+\rho ^{2\kappa})^{2}]$. Inflection points of
$U^{+}(l)$ imply peaks in
$C^{+}_{l}$ and allow to identify the
critical angular momenta beyond which some ``scattering phase transitions"
(quasi-bound states) would occur. As a matter of fact, this was the standpoint
in \cite{dos} leading to DO quasi-bound
states starting with the critical angular momentum for the MFE lens
$l_{cr}=6.876$.

\bigskip

{\em (iii) Closeness to reflectionlessness:}
Finally, let us mention that if one uses the Langer change of variable
($\rho=e^{x}$) and
function ($\psi =e^{x/2}\phi$) in the MFE nodeless
sector \cite{dosr}, the MFE focusing potential turns into the
following one $V_{1}(x)=-(n-1/2)(n+1/2)/{\rm cosh}^2x$.
Thus, the MFE case, and in general the DO class of focusing potentials are
not reflectionless, a property
requiring $n(n+1)$ in the numerator \cite{cr}. However, from their form,
one can claim that the DO
potentials are among the closest to the reflectionless ones. They miss
that quality by a ``semiclassical" 1/2 contribution.

To this end, we hint on the possible connection of
the isospectral effect with the morphology-dependent resonances of the MFE lens
\cite{mdr} and also on the possibility of an experimental search of the effect since 
spherical gradient-index sphere lenses are already a technological reality \cite{Jap}.

\begin{center} ***  \end{center}

This work was partially supported by the CONACyT Projects 4868-E9406
and 3898P-E9608 and by PROMEP/97.


\newpage

\vskip 2ex
\centerline{
\epsfxsize=280pt
\epsfbox{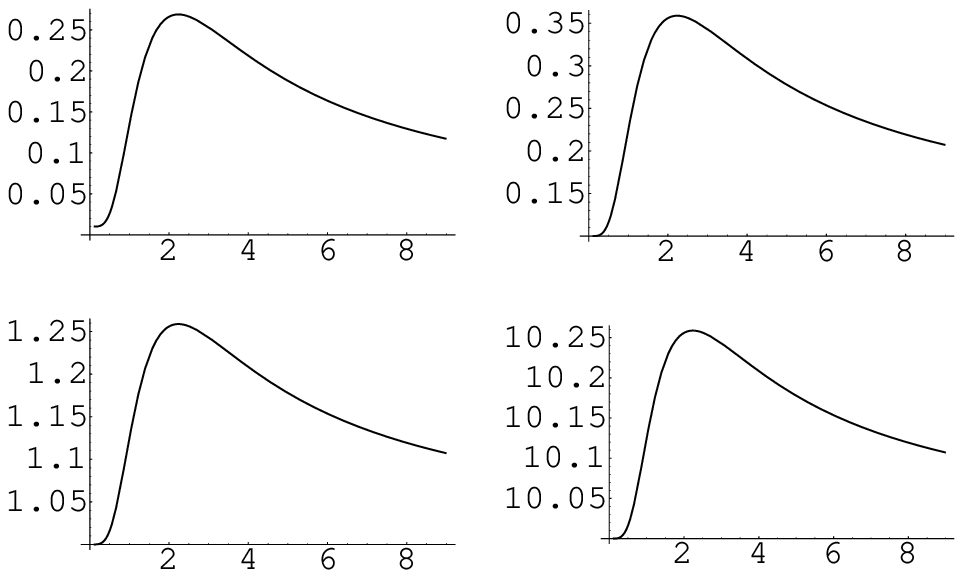}}
\vskip 4ex
\begin{center}
{\small{Fig. 1}\\
${\cal R} _{1}(\rho)$ for $\lambda=0.01, \;0.1,\;1,\;10$
}
\end{center}

\vskip 2ex
\centerline{
\epsfxsize=280pt
\epsfbox{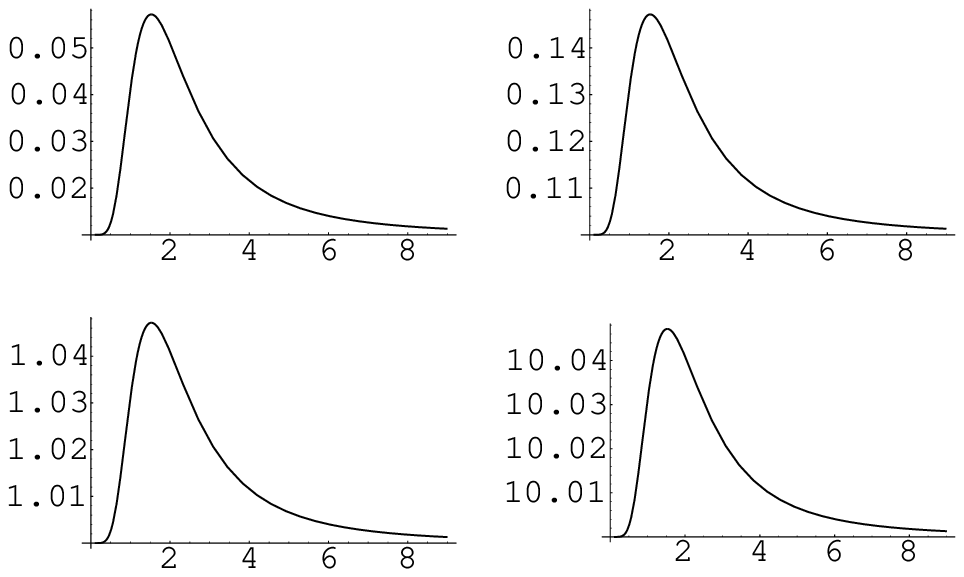}}
\vskip 4ex
\begin{center}
{\small{Fig. 2}\\
${\cal R} _{2}(\rho)$ for the same $\lambda$ values.
}
\end{center}

\vskip 2ex
\centerline{
\epsfxsize=280pt
\epsfbox{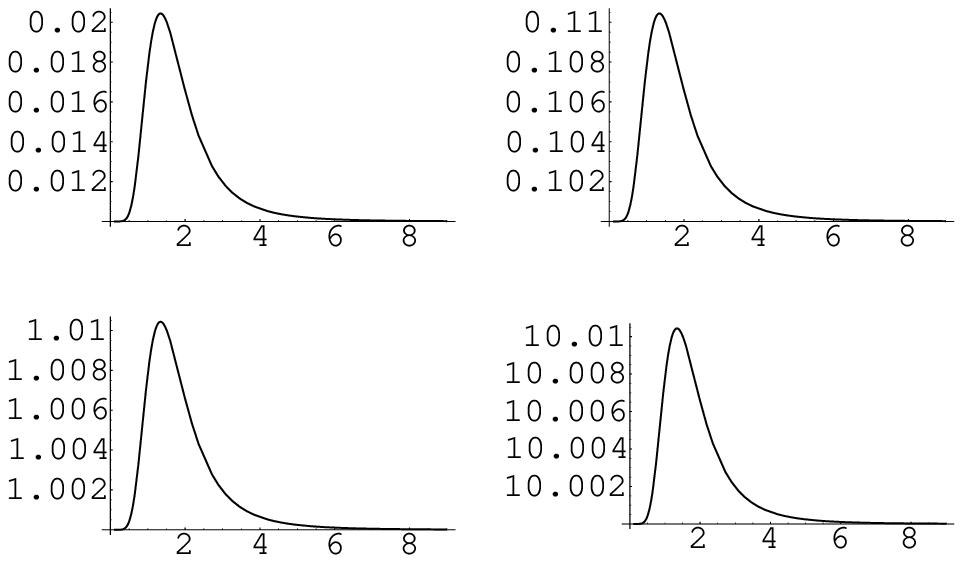}}
\vskip 4ex
\begin{center}
{\small{Fig. 3}\\
${\cal R} _{3}(\rho)$ for the same $\lambda$ values.
}
\end{center}

\end{document}